\documentclass[submission, PhysProc]{SciPost}

\usepackage{lscape,bm,amssymb,here}

\hypersetup{
    colorlinks,
    linkcolor={red!50!black},
    citecolor={blue!50!black},
    urlcolor={blue!80!black}
}

\DeclareSymbolFont{usualmathcal}{OMS}{cmsy}{m}{n}
\DeclareSymbolFontAlphabet{\mathcal}{usualmathcal}

\begin{document}

\begin{center}{\Large \textbf{
Electric-dipole transitions in $^6$Li with a fully microscopic six-body calculation
}}\end{center}

\begin{center}
W. Horiuchi\textsuperscript{$\star$} and
S. Satsuka
\end{center}

\begin{center}
Department of Physics, Hokkaido University, Sapporo 060-0810, Japan
\\
${}^\star$ {\small \sf whoriuchi@nucl.sci.hokudai.ac.jp}
\end{center}

\begin{center}
\today
\end{center}


\definecolor{palegray}{gray}{0.95}
\begin{center}
\colorbox{palegray}{
  \begin{tabular}{rr}
  \begin{minipage}{0.05\textwidth}
    \includegraphics[width=14mm]{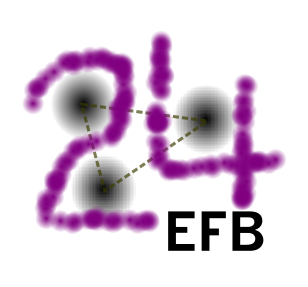}
  \end{minipage}
  &
  \begin{minipage}{0.82\textwidth}
    \begin{center}
    {\it Proceedings for the 24th edition of European Few Body Conference,}\\
    {\it Surrey, UK, 2-4 September 2019} \\
    \end{center}
  \end{minipage}
\end{tabular}
}
\end{center}

\section*{Abstract}
{\bf
Exploring new excitation modes 
and the role of the nuclear clustering has been of great interest. 
An interesting speculation 
 was made in the recent photoabsorption measurement of $^6$Li 
that implied the importance of the nuclear clustering.
To understand the excitation mechanism of $^6$Li,
we perform a fully microscopic six-body calculation on
 the electric-dipole ($E1$) transitions 
and discuss how $^6$Li is excited by the  $E1$ field as a function 
of the excitation energy. 
We show the various cluster components 
in the six-body wave functions
 and discuss the role of the nuclear clustering 
in the $E1$ excitations of $^6$Li.
}

\vspace{10pt}
\noindent\rule{\textwidth}{1pt}
\tableofcontents\thispagestyle{fancy}
\noindent\rule{\textwidth}{1pt}
\vspace{10pt}

\section{Introduction}
\label{sec:intro}

Nuclear cluster structure often appears in the spectrum of light nuclei. 
The role of the nuclear clustering in the electric-dipole ($E1$) transitions 
has been of great interest since they are closely related to 
important astrophysical reactions. 
Recently, an interesting speculation on the $E1$ excitation mechanism 
was made in the measurement of the photoabsorption cross section of $^6$Li
 that implied the coexistence of the typical and cluster $E1$ 
excitation modes~\cite{Yamagata17}. 
To understand this excitation mechanism,
we performed a fully microscopic six-body calculation 
with the correlated Gaussian method, 
in which the formation and distortion of the nuclear clusters 
are naturally taken into account as was demonstrated
for $^6$He~\cite{Mikami14}. 
Basically, this contribution aims at reviewing 
the discussions and findings given in Ref.~\cite{Satsuka19}
accompanying with some unpublished results.
We calculate the $E1$ transition strengths and their transition densities 
and discuss how $^6$Li is excited by the $E1$ field as a function 
of the excitation energy. 
We find the out-of-phase transitions due to
the valence proton and neutron
around the alpha cluster dominate in the low-lying energy regions 
below the alpha breaking threshold indicating “soft” 
Goldhaber-Teller (GT~\cite{Goldhaber48}) excitation, which 
is very unique in $^6$Li, whereas the typical GT mode appears 
in the higher-lying energy regions. 
We discuss the role of the nuclear clustering 
in the $E1$ excitations of $^6$Li by
showing the various cluster components 
in the six-body wave function.

In Sec.~\ref{method.sec}, we describe
the setup of the microscopic six-body calculation.
How we construct the six-nucleon wave functions are briefly explained.
Section~\ref{results.sec} presents calculated results and 
discusses the structure of the $E1$ excitations of $^6$Li
through the $E1$ transition strengths in Sec.~\ref{e1ts.sec} 
and its transition densities in Sec.~\ref{e1td.sec}.
In Sec.~\ref{discussion.sec}, the isoscalar dipole transitions
are discussed as complementary information of the ordinary $E1$ excitation.
Conclusion is made in Sec.~\ref{conclusion.sec}.

\section{Microscopic six-body calculation for $^{6}$Li}
\label{method.sec}

\subsection{Hamiltonian}

The Hamiltonian for a six-nucleon system is set by
the kinetic energy and two-body nuclear and Coulomb potential terms
in the center-of-mass (cm) frame. For the sake of simplicity,   
we employ the Minnesota (MN) nucleon-nucleon potential~\cite{MN} which
offers a fair description of the binding energies
and radii of $s$-shell nuclei, $^2$H ($d$), $^3$H ($t$), $^3$He ($h$), 
and $^4$He ($\alpha$)~\cite{Varga95, Suzuki08} without a three-body force.
The MN potential includes the one parameter $u$
responsible for the strength of odd-parity partial waves.
Roughly speaking, the $u$ parameter controls the interaction of
the valence nucleons from the $\alpha$ core on the $p$-shell orbital. 
Since the original strength ($u=1.00$) 
does not give the correct low-lying threshold and rms radius of $^6$Li,
the other two parameter sets are examined that are chosen to reproduce
the three-body threshold of $\alpha+p+n$ ($u=0.93$)
and the rms radius of $^6$Li ($u=0.87$), respectively.

\subsection{Correlated Gaussian expansion}

To describe the six-nucleon ground and excited states,
we expand the spatial part of the wave function in terms of 
the correlated Gaussian function
with the global vector representation~\cite{Varga95, SVM}:
$\exp\left(-\sum_{j,k=1}^5A_{jk}\bm{x}_j\cdot\bm{x}_k\right)
\mathcal{Y}_{LM_{L}}(
\sum_{j=1}^{5}v_j\bm{x}_j)$.
The coordinate set $(\bm{x}_1,\dots,\bm{x}_{5})$ 
excluding the cm coordinate of the six-nucleon system, $\bm{x}_6$,
is taken as the Jacobi coordinate.
The correlations among the particles
are explicitly included through the off-diagonal parameter,
$A_{jk}$ ($j\neq k$).
The rotational motion of the system
is described with the so-called global vector
in the argument of the solid spherical harmonic~\cite{Suzuki98,SVM}.
This expression is convenient that
the functional form does not change under
any linear transformation of the coordinate.
Various configurations such as 
single-particle, $\alpha+p+n$ and $h+t$ cluster configurations
can easily be implemented.
The matrix elements of the Hamiltonian can be evaluated analytically.
See~\cite{Varga95, SVM, Suzuki08} for 
the detailed mathematical derivation and expressions.
With these nice property,
the correlated Gaussian has been applied to 
studying the nuclear clustering~\cite{Horiuchi08, Horiuchi14, Ohnishi17}.
See, also review papers~\cite{Mitroy13, Suzuki17}.
The six-nucleon spin and isospin functions as well as antisymmetrization
of the total basis function are fully taken into consideration.
The isospin mixing due to the Coulomb interaction is naturally
described in this study.

\subsubsection{Ground-state wave function}

\begin{figure}[H]
\centering
\includegraphics[width=7cm]{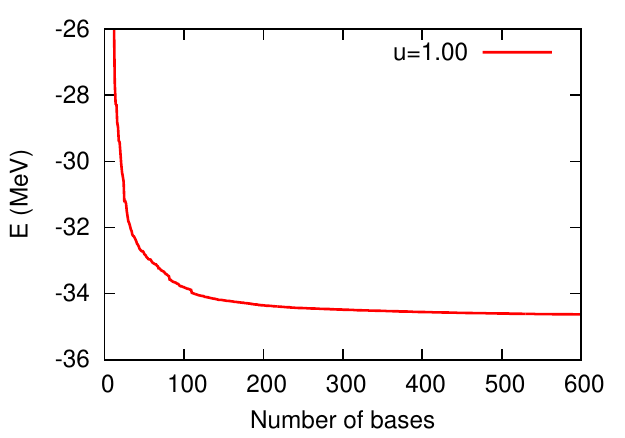}
\caption{Energy convergence of $^6$Li as a function of the number of bases.}
\label{convergence.fig}
\end{figure}

To find an optimal choice of a huge number of the variational parameters,
we employ the stochastic variational method  (SVM)~\cite{Varga95,SVM}.
We increase the number of basis selected competitively
from randomly generated candidates
until a certain number of basis states are obtained.
Figure~\ref{convergence.fig} displays the obtained energy curve 
 with $u=1.00$ as a function of the number of bases.
We see the energy of the six-nucleon system is rapidly converged
with increasing the number of the basis functions.
We stop the calculation with 600 bases,
which are very small by noting that
each basis function includes $6(6-1)/2=15$ parameters
as well as the spin degrees of freedom.
Then we switch the selection procedure
for the refinement of the variational parameters in
the already obtained basis functions
until the energy is converged within tens of keV.
For the wave functions with other $u$ parameters,
we start with the optimal basis functions with $u=1.00$ and
refine those basis functions by keeping the total number of
basis unchanged until the energy convergence is reached.

\subsubsection{Construction of excited states}

\begin{figure}[H]
\begin{center}
\includegraphics[width=12cm]{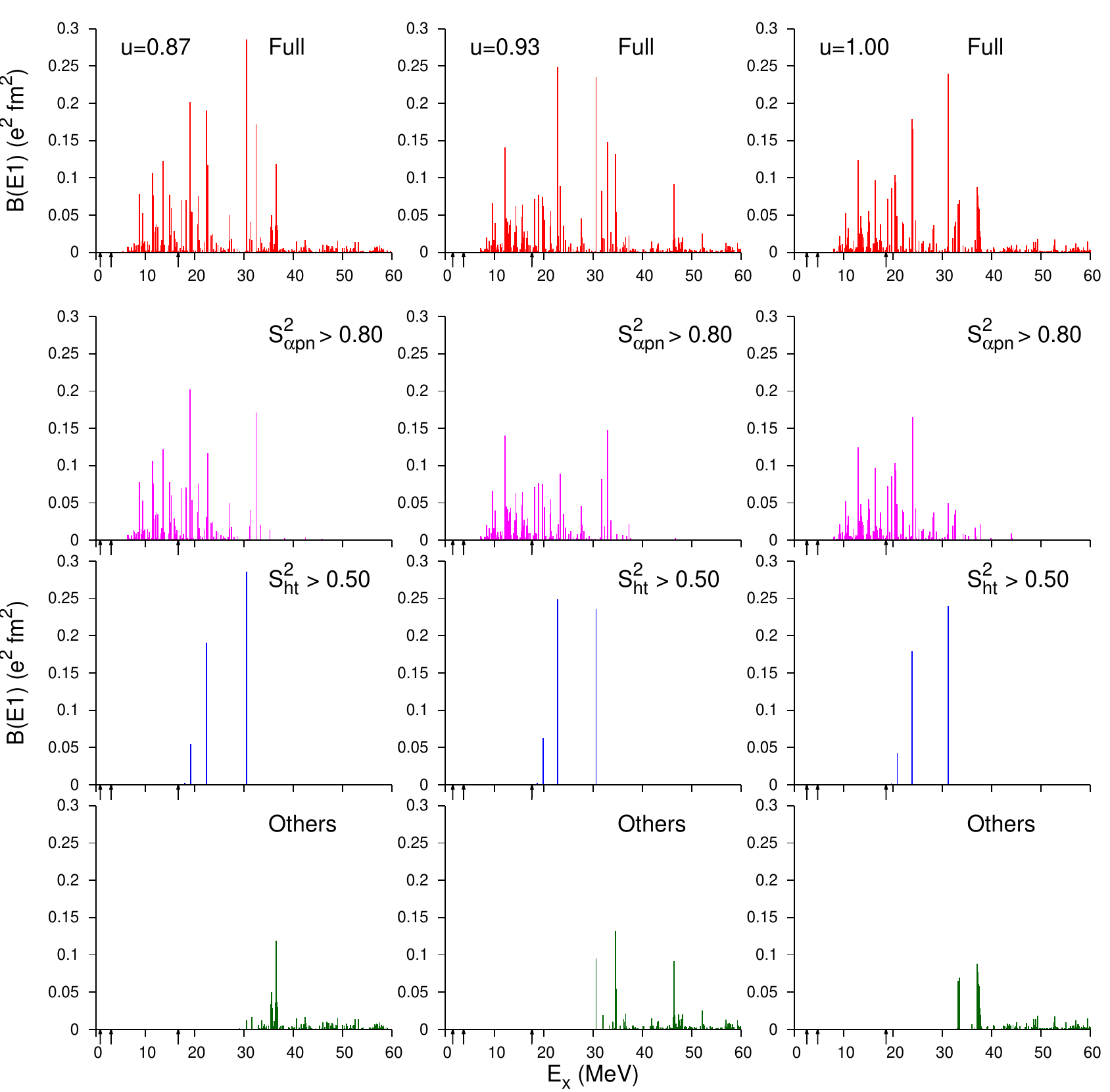}
\caption{Electric-dipole transition strengths and their decomposition with
  respect to the spectroscopic factors. The parameters
  $u=0.87, 0.93$ and 1.00 in the MN interaction are employed.
  Arrows indicate the theoretical threshold energies of $\alpha+d$,
  $\alpha+p+n$, and $h+t$ from left to right, respectively.
  The results with $u=0.93$ and the Full calculations
  are adopted from Ref.~\cite{Satsuka19}.}
\label{e1str.fig}
\end{center}
\end{figure}

Here we overview how to construct 
the final-state wave functions excited by the $E1$ operator:
$\mathcal{M}(E1,\mu)=e\sqrt{\frac{4\pi}{3}}\sum_{i\in p}\mathcal{Y}_{1\mu}
(\bm{r}_i-\bm{x}_6)$,
where $\bm{r}_i$ is the single-particle coordinate of the $i$th proton.
All the details are described in Ref.~\cite{Satsuka19}. 
We expand the final-state wave function in a large number
of the correlated Gaussian basis functions.
To incorporate the six-body correlations efficiently,
physically important configurations are selected:
(I) Single-particle excitation, (II) 4+1+1 cluster, 
and (III) 3+3 cluster configurations.

The configurations of type (I) is based on the idea that
the $E1$ operator excites one coordinate in the ground-state wave function.
The resulting coherent states are important
to satisfy the $E1$ sumrule~\cite{Horiuchi12,Horiuchi13,Mikami14}.
They are constructed by
using the basis set of the ground-state wave function of $^6$Li by multiplying
an additional solid spherical harmonic with the angular momentum $1$. 
The configurations of types (II) and (III) explicitly describe
the cluster configurations of $\alpha+p+n$ and $h+t$,
which correspond to the two lowest thresholds.
The relative wave functions of the valence nucleons
are expanded with several Gaussian functions
covering from short to far distances up to about 20 fm.

We include all the basis states for each subsystem independently.
Therefore, the final-state wave functions are not restricted
to the subsystems being the ground state but
the excitations and distortion of $^6$Li, $\alpha$, $h$, and $t$ are
included through the coupling of the pseudo excited states of those
nuclear systems.
We diagonalize the Hamiltonian including all the configurations
of types (I)--(III)
with 18490 basis functions and find $\sim 2\times 10^3$ states below
the excitation energy of 100 MeV.

Though we employed the large model space mentioned above,
the states are discretized since the wave functions
are expanded by square-integrable bases.
For more quantitative discussions,
it is necessary to include the continuum effect explicitly
by employing, e.g., the complex scaling method~\cite{CSM}
and Lorentz integral transform method~\cite{LIT}.
However, this is involved and the beyond the scope
of the present analysis.

\section{Results and discussions}
\label{results.sec}

\subsection{Electric-dipole transition strengths} 
\label{e1ts.sec}

\begin{table}[H]
\begin{center}
  \caption{Excitation energy $E_x$ in MeV,
    $B(E1)$ in $e^2$fm$^2$ and $\alpha+p+n$
    and $h+t$ spectroscopic factors in $^6$Li
    with different $u$ parameters.
    The results with $u=0.93$ are adopted from
Ref.~\cite{Satsuka19}.}
\footnotesize
    \begin{tabular}{cccccccccccccccc}
      \hline\hline
      \multicolumn{4}{c}{$u=0.87$}&&\multicolumn{4}{c}{$u=0.93$}&&\multicolumn{4}{c}{$u=1.00$}\\
\cline{1-4}\cline{6-9}\cline{11-14}      
$E_x$&$B(E1)$&$S_{\alpha pn}^2$&$S_{ht}^2$&&$E_x$&$B(E1)$&$S_{\alpha pn}^2$
&$S_{ht}^2$&&$E_x$&$B(E1)$&$S_{\alpha pn}^2$&$S_{ht}^2$\\
\hline
 8.9&0.076&1.000&0.000&& 9.6&0.066&0.999&0.000&&10.4&0.052&0.999&0.001\\
11.5&0.106&0.999&0.006&&12.1&0.140&0.999&0.011&&12.9&0.124&0.999&0.009\\
11.7&0.076&0.999&0.007&&14.3&0.063&0.997&0.000&&15.1&0.055&0.997&0.000\\
13.6&0.122&0.997&0.000&&15.7&0.064&0.999&0.010&&16.3&0.097&0.998&0.016\\
\\
19.1&0.202&0.988&0.005&&18.9&0.077&0.991&0.004&&20.3&0.103&0.993&0.001\\
20.8&0.076&0.994&0.010&&19.8&0.075&0.994&0.003&&20.5&0.094&0.995&0.001\\
22.4&0.191&0.124&0.845&&22.8&0.249&0.113&0.850&&23.8&0.179&0.133&0.827\\
22.7&0.117&0.948&0.013&&23.3&0.089&0.963&0.003&&23.9&0.165&0.880&0.085\\
\\
30.6&0.285&0.035&0.691&&30.6&0.236&0.264&0.533&&31.2&0.240&0.162&0.618\\
32.5&0.172&0.918&0.005&&30.6&0.095&0.780&0.158&&33.5&0.070&0.712&0.002\\
35.7&0.050&0.594&0.002&&33.0&0.148&0.962&0.005&&37.1&0.088&0.045&0.001\\
36.6&0.119&0.351&0.002&&34.6&0.132&0.195&0.019&&37.2&0.077&0.052&0.003\\
\hline\hline
  \end{tabular}
  \label{spect.tab}
\end{center}
\end{table}

Figure~\ref{e1str.fig} plots
the $E1$ transition strengths
or reduced transition probability $[B(E1)]$
obtained with the full model space
 that includes the configurations of types (I)--(III)
 with different values of the
$u$ parameter as a function of the excitation energy, $E_x$.
Large $B(E1)$ values in the low($E_x\lesssim 16$ MeV)-,
intermediate($E_x\sim 16$--30 MeV)-,
and high($E_x\gtrsim 30$ MeV)-energy regions are found.
There is little quantitative difference among
these three different $u$ values up to $E_x\sim 40$ MeV.

The decomposition of the $E1$ transition strengths
with respect to the spectroscopic factors
of $\alpha+p+n$ ($S_{\alpha pn}^2$) and $h+t$ ($S_{ht}^2$)
are also plotted in Fig.~\ref{e1str.fig}.
Here $S_{\alpha pn}^2$ and $S_{ab}^2$ are respectively defined
by
 $\left|\left<\Psi^{(\alpha)}\Psi^{(p)}\Psi^{(n)}\right|\left.\Psi^{(6)}\right>\right|^2$
 and
 $\left|\left<\Psi^{(a)}\Psi^{(b)}\right|\left.\Psi^{(6)}\right>\right|^2$
 with the wave function $\Psi^{(X)}$ of $X(=p, n, d, h, t, \alpha)$ cluster
 and $X=6$ denoting the six-nucleon system.
 In the ground state of $^6$Li, the $S_{\alpha d}^2$ value
 is large $\sim 0.9$~\cite{Satsuka19} for all the $u$ parameters employed,
 indicating a well-developed $\alpha+d$ cluster structure.
As we see in the $E1$ transitions to the 
states with $S_{\alpha pn}^2 >0.80$, 
most of the low-lying states below 20 MeV have
a large $\alpha+p+n$ cluster component.
The $E1$ strengths
with $S_{ht}^2>0.50$ show three large $E1$ strengths
after the $h+t$ threshold 15.8 MeV~\cite{Tilley02}, opens.
These robust structures do not depend on the $u$ parameter.
The first two can be the observed
levels at $E_x=17.98$ and 26.59 MeV, having relatively small
$h+t$ decay widths, 3.0 and 8.7 MeV for the first and second peaks,
respectively~\cite{Tilley02}.
The other complementary strengths are plotted in the lowest three panels of
Fig.~\ref{e1str.fig}. Since all the particle thresholds are open
beyond 30 MeV, many small strengths are distributed
due to the coupling of various configurations

For quantitative discussions, Table~\ref{spect.tab} summarizes
$E_x$, $B(E1)$, $S_{\alpha pn}^2$, and $S_{ht}^2$
of the states that give four largest $B(E1)$ values
in the low-, intermediate-, and high-energy regions 
with $u= 0.87, 0.93$, and 1.00. 
We note that in the low energy regions ($E_x\lesssim$16 MeV) below
the $h+t$ threshold, all the states have large
$S_{\alpha pn}^2$ values being $\sim 1$,
whereas their $S_{ht}^2$ values are $\sim 0$.

\begin{figure}[H]
\begin{center}
\includegraphics[width=9cm]{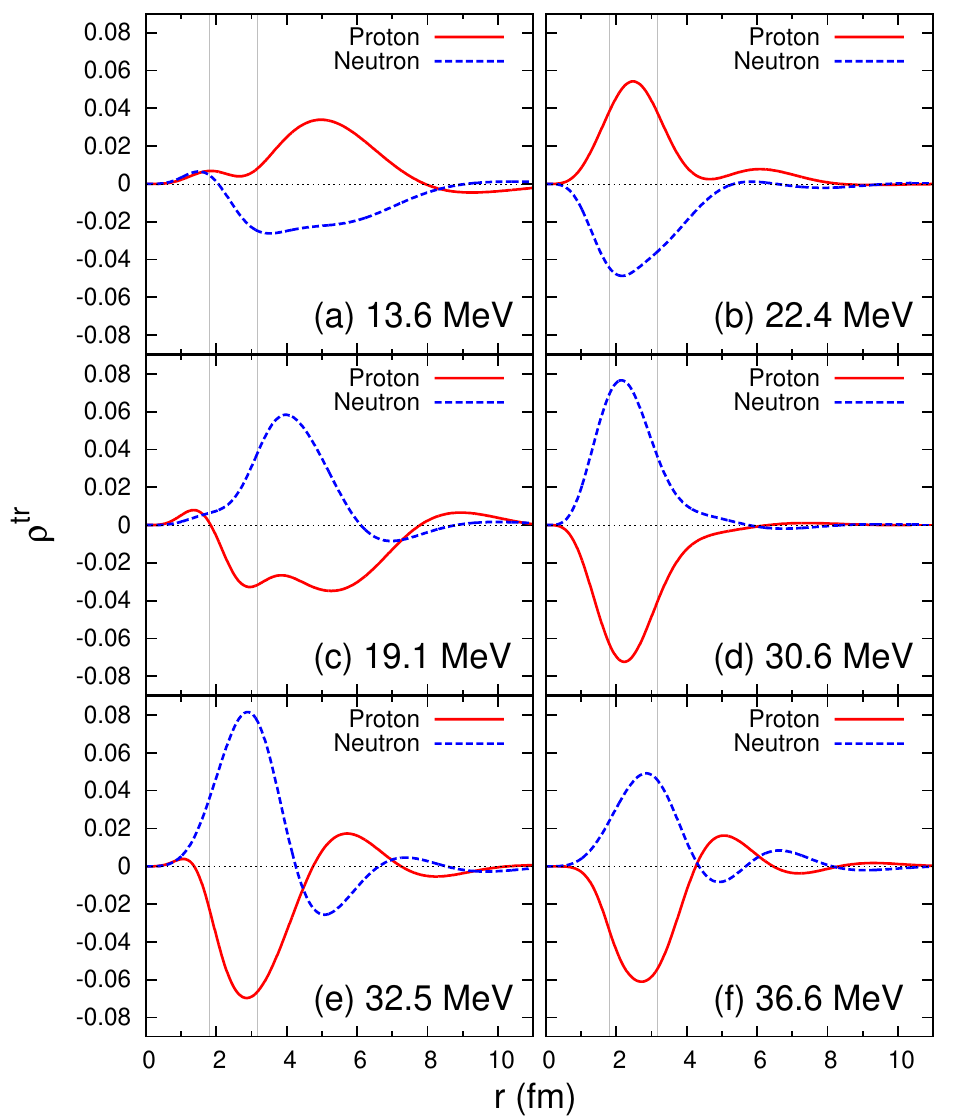}
\caption{Transition densities for proton and neutron with
  $u=0.87$ at (a) 13.6, (b) 22.4, (c) 19.1, (d) 30.6, (e) 32.5,
  and (f) 36.6 MeV.
Vertical thin lines indicate theoretical nuclear radii,
$^4$He and $^6$Li, respectively.}
\label{trdens.fig}
\end{center}
\end{figure}

\subsection{Dipole transition densities}
\label{e1td.sec}

To discuss the structure of the $E1$ excitations,
here we calculate the transition densities:
$\rho^{\rm tr}_{p/n}(r)=
\sum_{i\in p/n}\left<J_f\right\|
\mathcal{Y}_{1}(\bm{r}_i-\bm{x}_{6})\delta(|\bm{r}_i-\bm{x}_6|-r)
\left\|J_0\right>$, where $J_0$ and $J_f$ are the angular momentum
of the ground and final states, respectively.
Note that the $E1$ transition matrix can be obtained with
$\left<J_f\right\|\mathcal{M}(E1)\left\|J_0\right>=$
$e\sqrt\frac{4\pi}{3}\int_0^\infty dr\,\rho^{\rm tr}_{p}(r)$,
which express the spatial distribution of the $E1$ transitions.
In the following, we discuss the transition densities
for $^{6}$Li of the selected transitions
that show some characteristic behaviors.

\subsubsection{Soft Goldhaber-Teller excitations}

Figure~\ref{trdens.fig} displays $\rho^{\rm tr}_{p}$ and $\rho^{\rm tr}_{n}$
to the states that have the prominent $B(E1)$ values.
We discuss the results with $u=0.87$, which were not presented
in Ref.~\cite{Satsuka19}, though they are qualitatively the same
as those obtained with $u=0.93$.
The transition densities such at (a) $E_x=13.6$, (c) 19.1, and (e) 32.5 MeV
can be categorized into the same group.
All the transition densities show the in-phase transition
around the $^4$He radius and the out-of-phase transitions
beyond the nuclear surface.
More oscillations in the outside region
appear with increasing the excitation energy.
We interpreted this unique transition as
``soft'' Goldhaber-Teller(GT)-dipole mode, that is,
the oscillation of the valence proton and neutron
around the core ($\alpha$)~\cite{Satsuka19},
which is a variant of the classical picture of the
giant dipole resonance (GDR)~\cite{Goldhaber48}.
In Table~\ref{spect.tab}, we see all these states
a large $S_{\alpha pn}^2$ value $\gtrsim$ 0.9.
It should be noted that this mode is different
from the so-called soft-dipole mode
as Ref.~\cite{Mikami14} showing the oscillation of
the valence two neutrons against the core~\cite{Mikami14}.

\subsubsection{Cluster and Goldhaber-Teller excitations}

At (b) $E_x=22.4$, (d) $30.6$, and (f) $36.6$ MeV,
all the transition densities exhibit
out-of-phase transitions in all regions,
showing the typical GT mode.

Let us first discuss the states at (b) $E_x=22.4$ and (d) $30.6$ MeV.
In this energy region, the $h+t$ threshold already opens.
Considering the fact that the $S_{ht}^2$ values are large $\gtrsim 0.7$
(see Table~\ref{spect.tab}),
this behavior can be interpreted as the $E1$ excitation of
the relative wave function between the $h$ and $t$ clusters.
Actually, the peak positions are located at $\sim$2 fm,
which is about the sum of the peak positions
of the densities of $h$ and $t$~\cite{Satsuka19}.

We also see the state with the state with (f) $E_x=36.6$ MeV
also show the out-of-phase transition in all regions
but the $E1$ excitation structure is found being
different from those of (b) and (d).
The peak positions are located somewhat outside
from these of (b) and (d), and
the $S_{\alpha pn}^2$ and $S_{ht}^2$ are small
as listed in Table~\ref{spect.tab}.
This is nothing but the typical GT mode,
in which the protons and neutrons
oscillate against each other~\cite{Goldhaber48}.

\subsection{Discussion: Isoscalar dipole transitions}
\label{discussion.sec}

\begin{figure}[H]
\begin{center}
\includegraphics[width=13cm]{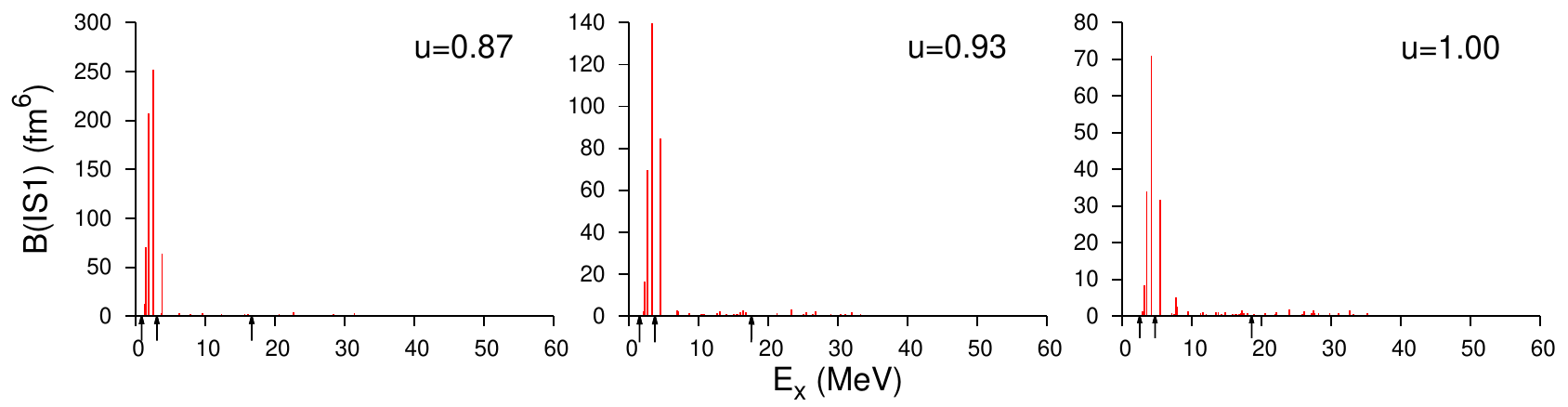}
\caption{Isoscalar dipole transition strengths 
  with $u=0.87$, 0.93, and 1.00.
  The results with $u=0.93$
are adopted from the data given in Ref.~\cite{Satsuka19}.}
\label{dipolestr.fig}
\end{center}
\end{figure}

\begin{figure}[H]
\begin{center}
\includegraphics[width=7cm]{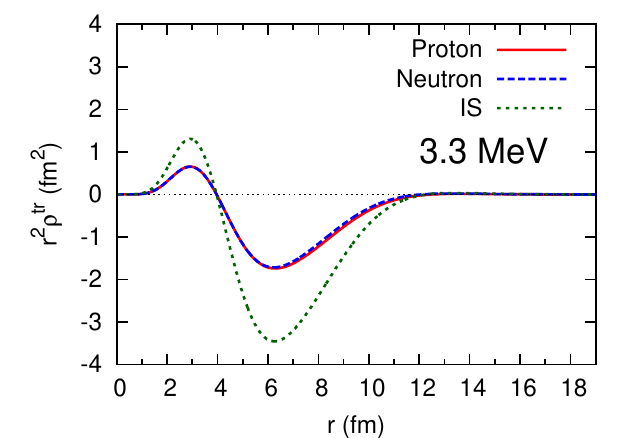}
\caption{Transition densities for proton and neutron
  and their sum $\rho_p^{\rm tr}+\rho_n^{\rm tr}$ (IS) multiplied by $r^2$
with $u=0.93$ at $E_x=3.3$ MeV.}
\label{trdensIS.fig}
\end{center}
\end{figure}

To discuss more details on the transition densities,
here we discuss the isoscalar dipole excitation whose operator
is defined by $\mathcal{M}({\rm IS}1,\mu)
=\sum_i(\bm{r}_i-\bm{x}_6)^2
\mathcal{Y}_{1\mu}(\bm{r}_i-\bm{x}_6)$~\cite{Stringari82}.
By definition, the IS1 transition matrix 
can be obtained with
$\int_{0}^\infty dr\, r^2 \left(\rho_{p}^{\rm tr}+\rho_{n}^{\rm tr}\right)$.

\begin{figure}[H]
\begin{center}
\includegraphics[width=9cm]{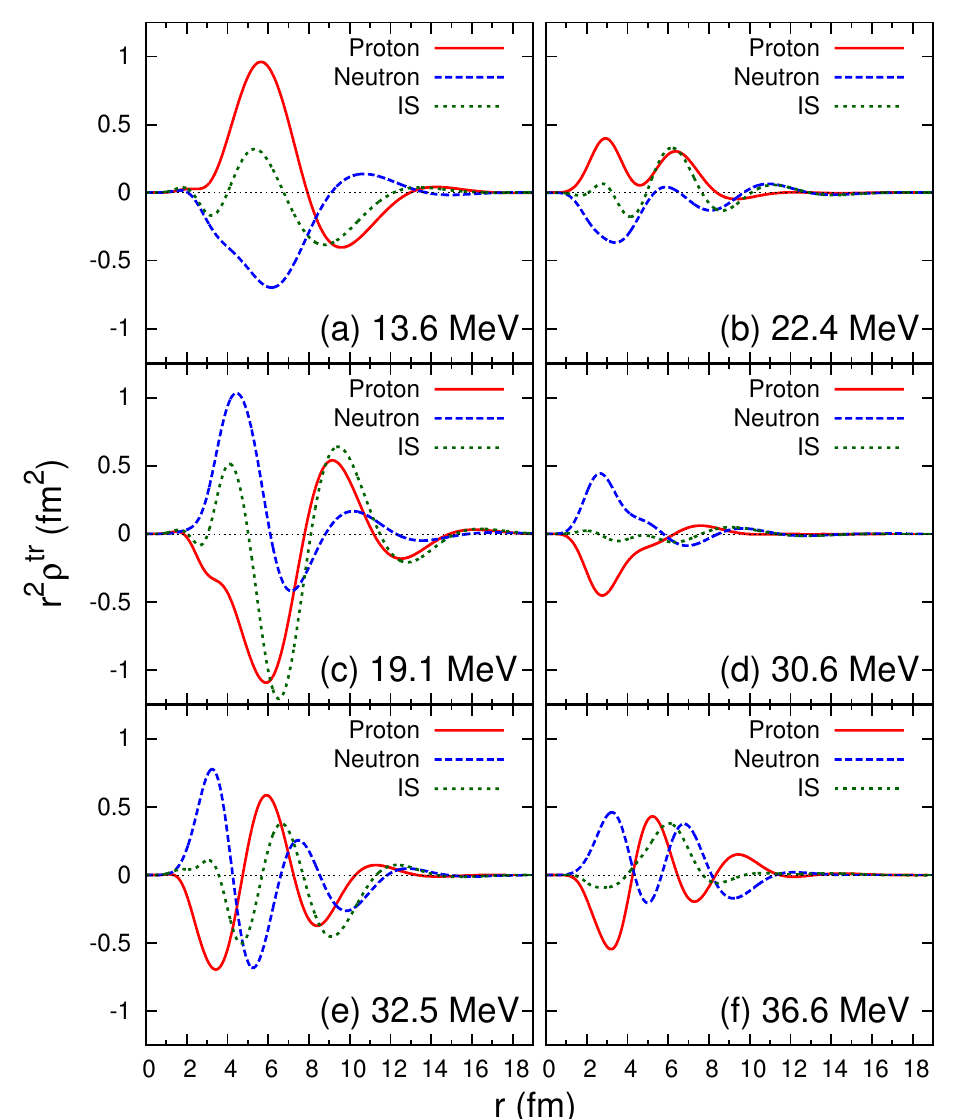}
\caption{Same as Fig.~\ref{trdensIS.fig}
  but with $u=0.87$ at the same $E_x$ of Fig.~\ref{trdens.fig}.}
\label{trdens-comp.fig}
\end{center}
\end{figure}

Figure~\ref{dipolestr.fig} displays the IS1 transition strengths
with different values of the $u$ parameter
as a function of the excitation energies. 
Some prominent strengths appear between
the $\alpha+d$ and $\alpha+p+n$ thresholds
and the $S_{\alpha d}^2$ values of those states are $\sim 1$.
They come from the transition to the $\alpha+d$ continuum states,
which cannot be excited by the ordinary isovector $E1$ operator.
Note that the isospin mixing components due to the Coulomb interaction
are included in the wave functions but they are small.
Figure~\ref{trdensIS.fig} draws
the transition densities multiplied by $r^2$ that show
the most prominent $B(IS1)$ value at $E_x=3.3$ MeV with $u=0.93$.
The transition densities of proton and neutron coincide each other,
showing the in-phase transition in all regions.

Let us see Fig.~\ref{trdens.fig} from a different view point.
Figure~\ref{trdens-comp.fig}
plots the transition densities of proton and neutron
multiplied by $r^2$ and their sum of the states given in Fig.~\ref{trdens.fig}.
We remind that the IS1 transition matrix is obtained by the sum
of the proton and neutron transition densities.
When the transition densities exhibit the out-of-phase transitions,
they are canceled, resulting in a small IS1 transition matrix.
In general, the in-phase transition enhances the IS1 transition matrix.
In the case of the soft GT-dipole mode,
since the in-phase transition occurs in the short distances,
the contributions from these regions 
become small due to the additional $r^2$ factor through the IS1 operator.
Since almost all the IS1 strengths are exhausted in
the low-energy regions below $\sim 5$ MeV,
only small strengths are found in the higher energy regions.
To understand the excitation mechanism of $^{6}$Li,
one can observe this strong suppression of the IS1 transition strengths
using such as $(\alpha,\alpha')$ measurement
as a complementary evidence of the existence of the soft-GT-dipole mode.


\section{Conclusion}
\label{conclusion.sec}

We performed fully microscopic six-body calculations
to understand the electric-dipole ($E1$) excitation mechanism.
The six-body wave functions were constructed in terms
of the correlated Gaussian (CG) functions.
The ground state wave function was obtained precisely
with the aid of the stochastic variational method.
The final state wave functions were also expressed by a number
of the CG functions. The asymptotic wave functions
between clusters as well as the distortion of the
clusters were taken into account.

We calculate the $E1$ transition strengths and
their transition densities.
With the analysis of the cluster components of
the wave functions,
we see that the nuclear clustering emerges in 
the $E1$ excitation depending on the positions
of the threshold energies.
We find very unique excitation mode,
``soft'' Goldhaber-Teller(GT) excitation,
which is recognized as the out-of-phase oscillation between
valence nucleons around the $\alpha$ cluster in $^6$Li,
dominates below the $\alpha$ breaking threshold energy.
Series of the vibrational excitation mode are also found
in the high-lying energy regions.

Beyond the $h+t$ threshold, the $E1$ transitions with
the $h+t$ cluster mode appear.
The transition densities show the out-of-phase transition in all regions.
With increasing the excitation energy
after all decay channels open beyond $\sim 30$ MeV,
the typical GT mode appears with largely distorted
configurations neither $\alpha+p+n$ nor $h+t$.

Isoscalar dipole (IS1) excitations are examined as it gives
a different view of the $E1$ excitation mechanism.
The out-of-phase transitions in the soft GT dipole mode
cause the strong suppression of the IS1 transition strengths.
A measurement of the $\alpha$ inelastic scattering cross sections
at around $\alpha+p+n$ may reveal the absence of the isoscalar
component at this energy regions, which will be a complementary
evidence of the existence of the soft GT dipole mode.
Also, exploring the appearance of the soft GT mode
in heavier nuclei is interesting, e.g., $^{18}$F as $^{16}{\rm O}+p+n$.

\section*{Acknowledgements}
 W. H. acknowledges the collaborative research program 2019,
information initiative center, Hokkaido University.


\paragraph{Funding information}
This work was in part supported by JSPS
KAKENHI Grants No. 18K03635, No. 18H04569, and No.
19H05140.

%






\end{document}